\documentclass{article}
\usepackage{graphicx}  
\usepackage{amsmath}   
\usepackage[compress,numbers,sort]{natbib}
\usepackage{amssymb}   
\usepackage{bm} 
\usepackage{dcolumn}
\usepackage{color}
\usepackage{mathrsfs}
\usepackage{amsfonts}
\usepackage{varioref}
\usepackage{textcomp}
\usepackage{mathtools}
\usepackage[normalem]{ulem}

\usepackage{multirow}
\usepackage{caption}
\usepackage{subcaption}
\RequirePackage[colorlinks,citecolor=blue,urlcolor=magenta,linkcolor=blue]{hyperref}
\allowdisplaybreaks
\labelformat{section}{Section #1} 
\labelformat{subsection}{Section #1} 
\labelformat{subsubsection}{Section #1}
\labelformat{subsubsubsection}{Section #1}
\labelformat{equation}{Eq.~(#1)} 
\labelformat{figure}{Fig.~#1} 
\labelformat{subfigure}{Fig.~\thefigure#1} 
\labelformat{table}{Table~#1} 
\labelformat{appendix}{Appendix #1}

\title{Tidal Love numbers and quasi-normal modes of the
ECO in a Dark Matter halo}
\author{ Kabir Chakravarti \footnote{ chakravarti@fzu.cz}$~^{1}$ and Chiranjeeb Singha\footnote{chiranjeeb.singha@iucaa.in}$~^{2}$\\
$^{1}$\small{CEICO, FZU-Institute of Physics of the Czech Academy of Sciences,  Na Slovance 1999/2,}\\
\small{ 182 21 Prague 8, Czech Republic}\\
$^{2}$\small{Inter-University Centre for Astronomy \& Astrophysics, Post Bag 4, Pune 411 007, India}
}

\begin{document}
  
\maketitle
\begin{abstract} 
It is well-known that exotic compact objects (ECOs) are a class of objects categorised as Black Hole (BH) mimickers. ECOs have been shown to possess signatures distinguishing them from BHs. However, in our universe, no object exists in complete isolation. Consequently, any compact object, whether a BH or an ECO, must reside within some environment that inevitably influences the surrounding spacetime geometry due to back-reaction. In this paper, we investigate a scenario where an ECO is embedded in an environment of dark matter (DM). In this work, we assume two different models of the DM halo profile. We compute the Love numbers and GW echoes of this composite system to assess the impact of the surrounding dark matter halo. To analyze the echoes, we focus on odd-parity perturbations, while for calculating the tidal Love numbers, we consider both even and odd parity perturbations. We aim to understand how the DM properties couple to the ECO signatures in the Love number or the GW-echo signal, both of which have a strong bearing as observables in future-generation detectors.

\end{abstract}

\section{Introduction}\label{sec:intro}

The observation of gravitational waves (GW) from merging compact objects, such as black holes (BHs),  and neutron stars, has opened a new observational window into the universe and initiated the phenomenological exploration of gravity in its highly dynamical regime \cite{LIGOScientific:2016aoc,KAGRA:2021vkt,LIGOScientific:2021sio}. Upcoming next-generation detectors are expected to significantly enhance the precision of these observations, enabling stringent tests of General Relativity (GR). These advancements will allow us to investigate a broad range of fundamental questions, including the validity of the Kerr hypothesis, potential deviations from GR, and extensions beyond the Standard Model \cite{ET:2019dnz, LISA:2024hlh, Evans:2021gyd}.

Gravitational waveforms have predominantly been computed within the vacuum GR framework, and in select cases, within alternative theories of gravity \cite{Blanchet:2013haa, Pretorius:2005gq, Berti:2015itd}. While such treatments are generally sufficient for detection, achieving unbiased parameter estimation and conducting high-precision tests of GR will likely require a comprehensive understanding of the effects introduced by surrounding matter environments. These effects must be appropriately modeled and incorporated into gravitational wave data analysis pipelines to ensure the reliability of scientific inferences \cite{LISA:2022kgy}. Substantial progress has been made in evaluating the influence of environmental matter on gravitational wave sources. While most of these efforts have considered standard baryonic components around compact objects, such as accretion disks or diffuse gas, some works have also considered dark matter (DM) media including exotic scenarios like axions \cite{Takahashi:2021yhy, Omiya:2024xlz}. An important step in this direction was taken by Barausse, Cardoso, and Pani \cite{Barausse:2014tra}, who carried out a systematic study of how astrophysical environments could affect gravitational wave signals. While such environments may act as sources of systematic uncertainties in tests of GR, they also present a valuable opportunity: gravitational waves can serve as probes of the environment itself, enabling astrophysical inference beyond the properties of the binary components. Among the various environmental scenarios, dark matter-dominated surroundings have emerged as particularly intriguing. These environments exhibit distinct physical signatures strongly dependent on the nature and interactions of the underlying dark sector. As such, they offer a promising avenue for constraining or unveiling new physics beyond the Standard Model. In the literature, a wide range of Newtonian and post-Newtonian models have been employed to capture the leading-order corrections to gravitational waveforms due to such environments, providing useful order-of-magnitude estimates and qualitative insights \cite{Eda:2013gg, Cardoso:2019rou, Hannuksela:2019vip}. However, as the sensitivity of gravitational wave detectors improves, these approximations may become insufficient. To match the precision capabilities of next-generation detectors, it becomes essential to develop fully relativistic models that faithfully incorporate the coupling between matter fields and the gravitational dynamics governed by GR. Motivated by these challenges, several studies over the past decade have focused on constructing relativistic frameworks that describe gravitational wave sources embedded in matter-rich environments \cite{Brito:2014wla,Pani:2015qhr,Macedo:2016wgh,Cardoso:2016oxy,Cardoso:2017cfl,
Cardoso:2017cqb,Hannuksela:2018izj,Berti:2019wnn,Cardoso:2019rvt,Cardoso:2019rou,Kavanagh:2020cfn,Coogan:2021uqv,Cardoso:2021wlq,Pereniguez:2021xcj,Speeney:2022ryg,DeLuca:2022xlz,Destounis:2022obl,Singh:2022wvw,Becker:2022wlo,DeLuca:2023mio,Katagiri:2023umb,Takahashi:2023flk,Figueiredo:2023gas,Traykova:2023qyv,Nichols:2023ufs,Brito:2023pyl,Destounis:2023ruj,Duque:2023seg,Speeney:2024mas,Santoro:2024tmo,Dyson:2024qrq,Chowdhury:2024auw,Mitra:2023sny,Bertone:2024rxe,Fischer:2024dte,Wilcox:2024sqs,Cannizzaro:2024yee,Cirelli:2024ssz,Rivera:2024xuv,Cannizzaro:2024hdg,Ianniccari:2024ysv,Spieksma:2024voy,Boyanov:2024jge, Berti:2024moe,DeLuca:2024ufn,DeLuca:2024uju,Cannizzaro:2024fpz,Tahelyani:2024cvk,Katagiri:2024wbg}. These include black hole binaries interacting with baryonic disks, compact objects within dense stellar clusters, and systems immersed in dark matter halos. In particular, considerable attention has been paid to the scenario in which particle dark matter forms dense halos, or ``spikes,'' around compact objects. This configuration was first proposed in the seminal work of Gondolo and Silk \cite{Gondolo:1999ef}, and was later extended to general relativistic settings \cite{Sadeghian:2013laa,Ferrer:2017xwm}. The influence of such dark matter spikes on the dynamics and waveform of inspiraling black hole binaries has been the subject of extensive investigation \cite{Eda:2013gg}. An initial estimate of waveform dephasing due to dynamical friction, relative to the vacuum GR case, was provided in \cite{Eda:2014kra}, and subsequent analyses \cite{Kavanagh:2020cfn,Coogan:2021uqv, Nichols:2023ufs} refined this picture by incorporating the dynamical response of the dark matter distribution. Dynamical friction effects have also been integrated into advanced waveform generation frameworks, such as the \texttt{FastEMRIWaveforms} package \cite{Speeney:2022ryg, Speeney:2024mas}, which aims to model extreme mass-ratio inspirals in complex environments. In parallel, significant efforts have been dedicated to developing a self-consistent and fully relativistic treatment of such systems by extending black hole perturbation theory to non-vacuum spacetimes \cite{Cardoso:2021wlq,Cardoso:2022whc}.

Recently, a fully relativistic model describing a Hernquist-type matter distribution surrounding a supermassive black hole was developed in~\cite{Chakraborty:2024gcr}. Motivated by this framework, the present work extends the analysis to a compact configuration where the central object is an exotic compact object (ECO) rather than a black hole. ECOs are proposed alternatives to classical BHs, arising in various quantum gravity-inspired models and modified theories of gravity \cite{Cardoso:2016rao,Cardoso:2016oxy,Cardoso:2017cqb, Mark:2017dnq,Correia:2018apm,Bueno:2017hyj,Abedi:2016hgu, Chakravarti:2021clm,Biswas:2023ofz}. A defining feature of ECOs is the absence of an event horizon, although they can closely resemble black holes in their external spacetime geometry. In this study, we construct a fully relativistic matter profile based on the Hernquist model and develop the linear perturbation theory around a static, spherically symmetric background supported by anisotropic matter. Our primary objective is to examine two key gravitational observables: quasinormal modes (QNMs) and tidal Love numbers (TLNs).
QNMs encapsulate the characteristic oscillation spectrum of the spacetime and can carry imprints of the surrounding matter, especially during the post-merger ringdown phase. To explore the potential presence of gravitational wave echoes and modifications to the ringdown signal, we focus on odd-parity (axial) perturbations. On the other hand, TLNs quantify the deformability of the central object under external tidal fields. We analyze both the even-parity (polar) and odd-parity (axial) sectors for their computation. By comparing our theoretical predictions with gravitational wave observations, this framework may offer novel avenues for constraining dark matter properties.

The paper is organized as follows: in \ref{Sec:Geo_ECO}, we describe the geometry of an ECO embedded in a dark matter halo. \ref{Sec:TIP_LN} is devoted to the computation of the tidal Love numbers for the ECO, considering both even and odd parity perturbations. In \ref{Sec:TDP_Echo}, we analyze the effects of time-dependent perturbations in the presence of a dark matter halo. Finally, our main findings are summarized in \ref{sec:conclusion}.

\textbf{\textit{Notations and Conventions}}:  
Throughout this paper, we adopt the mostly positive signature for the spacetime metric, where the Minkowski metric in $1+3$ dimensions takes the form $\text{diag}(-1, +1, +1, +1)$ in Cartesian coordinates. We work in geometrized units, as specified earlier.

\section{Geometry of an ECO in a dark matter halo}\label{Sec:Geo_ECO}
It is generally assumed in the literature that ECOs are stable structures of a yet undiscovered theory of gravity \cite{Cardoso:2016rao,Cardoso:2016oxy,Cardoso:2017cqb,Mark:2017dnq, Correia:2018apm,Bueno:2017hyj,Abedi:2016hgu,Chakravarti:2021clm, Biswas:2023ofz}. Consequently, the extension from a BH of mass $M$ to an ECO of the same mass under the assumption of spherical symmetry proceeds by replacing the BH horizon with an unknown surface. The ECO surface is parametrised by its reflectivity $\mathcal{R}$ and its location $r_\epsilon$. $r_\epsilon$ is usually modelled slightly shifted from the ECO's Schwarzschild Radius $r_S = 2M$ as
\begin{equation}\label{eq:r_epsilon}
    r_\epsilon = 2M(1 + \epsilon),
\end{equation}
here $\epsilon$ is a small parameter which tracks the deviation from classical behaviour. To implement the DM halo, we begin with a brief review of the literature. Our implementation of the DM around the ECO follows the development of \cite{Chakraborty:2024gcr}, which is itself based on the development of the topic presented in \cite{Cardoso:2021wlq}. We are interested ultimately in a physical picture where an ECO of mass $M$ is surrounded by a DM halo of mass $M_{\rm DM}$ and a scale-length $r_s$. To get to that, we start with the simple Hernquist type density profile \cite{1990ApJ...356..359H} for a DM halo whose radial behaviour is given by
\begin{align}\label{eq:rhox_H}
    4M^2\rho(x) &=  \frac{1}{2\pi}\left(\frac{x_{\rm DS}}{px}\right) (x + x_{\rm DS})^{-3}~, \nonumber \\
    \frac{m(x)}{2M} &= \frac{1}{p}\left(\frac{x}{x + x_{\rm DS}}\right)^2~.
\end{align}
Here $\rho(r)$ is the DM density and $m(r)$ is just the mass enclosed within a radial length $r$. Note that in \ref{eq:rhox_H}, we have rescaled the variables with factors of $M$ such that the equation is written exclusively in terms of dimensionless quantities. Accordingly, we introduce the reduced radial variable given by $x = r/2M$ as well as the reduced DM scale length $x_{\rm DS} = r_s/2M$. The quantity $p$ is defined as $p = 2M/M_{\rm DM}$. As is well known, \ref{eq:rhox_H} is the radial profile just with DM. When an object of mass $M$ is placed at $r=0$ in the DM halo, it undergoes an adiabatic contraction. This changes both the mass and the density profiles as given by Einstein's construction \cite{Cardoso:2021wlq} as,
\begin{align}\label{eq:rhox_NR}
    4M^2\rho(x) &=  \frac{1}{2\pi}\left(\frac{x_{\rm DS}+1}{px}\right) (x + x_{\rm DS})^{-3}\left(1-\frac{1}{x}\right) \nonumber~, \\
    \frac{m(x)}{2M} &= \frac{1}{2}+ \frac{1}{p}\left(\frac{x}{x + x_{\rm DS}}\right)^2\left(1-\frac{1}{x}\right)^2\nonumber~, \\
\end{align}
From \ref{eq:rhox_NR}, it is clear that in the absence of DM, $p\rightarrow\infty$ and hence the density vanishes and the mass just becomes equal to that of the ECO. In the presence of an object of mass $M$, it is readily seen that Einstein's construction implies anisotropic pressure, namely non-zero tangential pressure $P_t$ along with a vanishing radial pressure $P_r$. The energy momentum tensor then assumes the form $T_\nu^\mu = diag[-\rho, 0, P_t, P_t]$. Substituting this form for $T_\nu^\mu$ and then applying the Bianchi identity $\nabla_\mu T_\nu^\mu = 0$ implies
\begin{align}\label{eq:bianchi}
    \left[\frac{2P_t(x)}{\rho(x)}\right] = \frac{m(x)}{M x - 2m(x)} \nonumber \\
    4 \pi M^2 x^{2}\rho(x) = \frac{d}{dx}\left[\frac{m(x)}{2M}\right]
\end{align}
It should be noted that \ref{eq:bianchi} assumes the $\nabla$ operator is taken over a spherically symmetric background whose line element in terms of the reduced radial variable $x$ is given by
\begin{equation}\label{eq:metric_ansatz}
d\left(\frac{s}{2M}\right)^2=-e^{2N(x)} d\hat{t}^2+\frac{d x^2}{g(x)}+x^2 d\Omega^2~,
\end{equation}
where $\hat{t}= t/2M$ is the reduced time. \ref{eq:rhox_NR} is, however, based on a non-relativistic formulation of the DM halo. Very recently, this was extended to a fully relativistic formulation in \cite{Chakraborty:2024gcr}. In this formulation, the starting point is considered as the adiabatically contracted to a Hernquist-density profile, which gives rise to a spherically symmetric DM spike. Then the radial behaviour of the density is given by
\begin{equation}\label{eq:rho_SchHeq}
    4M^2\rho(x) = (4M^2\rho_0) \left[\left(1 - \frac{2}{x}\right)^w (2x)^{-q} \left(1 + \frac{x}{x_S^\prime}\right)^{q-4}\right],  
\end{equation}
where $\rho_0$ is a scaling constant having units of $1/M_\odot^2$. $w$ and $q$ are profile constants whose values are given by Table I of ref. \cite{Chakraborty:2024gcr}, and $x_S^\prime$ is a parameter given by  
\begin{equation}\label{eq:xs_prime}
    x_S^\prime = \frac{r_\mathrm{s}M_\mathrm{DM}}{M^2} = \frac{4 x_{DS}}{p}~.
\end{equation} 
It is readily seen that the density $\rho$ is divergent at $x = 2$, hence the DM spike is cutoff at $x\leq 2$. Then the form of the energy-momentum tensor is assumed to be exactly the same as that of the formulation for the non-relativistic case. The equations for $P_t(x)$ and $ m^\prime(x)$ are again given by \ref{eq:bianchi}. Integrating over the radial discontinuity in $\rho(x)$ gives us the following expression for $m(x)$ from \ref{eq:bianchi}   
\begin{equation}\label{eq:massx_NR}
m(x)=M+\lambda R_S^3 \tilde{C} \, _2F_1\left(w+1,q+w-2;w+2;-\frac{(x-2) x_s^\prime}{4( x+x_s^\prime)}\right)\Theta(x-2)~.
\end{equation}
To complete the geometry, we must also quantify the behaviour of the metric around such objects. This is straightforwardly achieved by substituting the metric ansatz of \ref{eq:metric_ansatz} into the field equations. We obtain the following useful relations for the metric potentials $N(r)$ and $g(r)$ from the $tt$ and the $xx$ components of the equations respectively
\begin{align}\label{eq:gN}
    g(x) &= 1-\frac{m(x)}{Mx} \nonumber \\
     \frac{dN(x)}{dx} &= \frac{ m(x)}{x[Mx-2m(x)]}~.
\end{align}
The actual integration for $N(x)$ in \ref{eq:gN} does not admit analytical expressions and has to be done numerically. We will make explicit use of the numerically integrated metric functions when calculating GW-echoes in \ref{Sec:TDP_Echo}. 
\begin{figure}[h]
    \includegraphics[width=\linewidth]{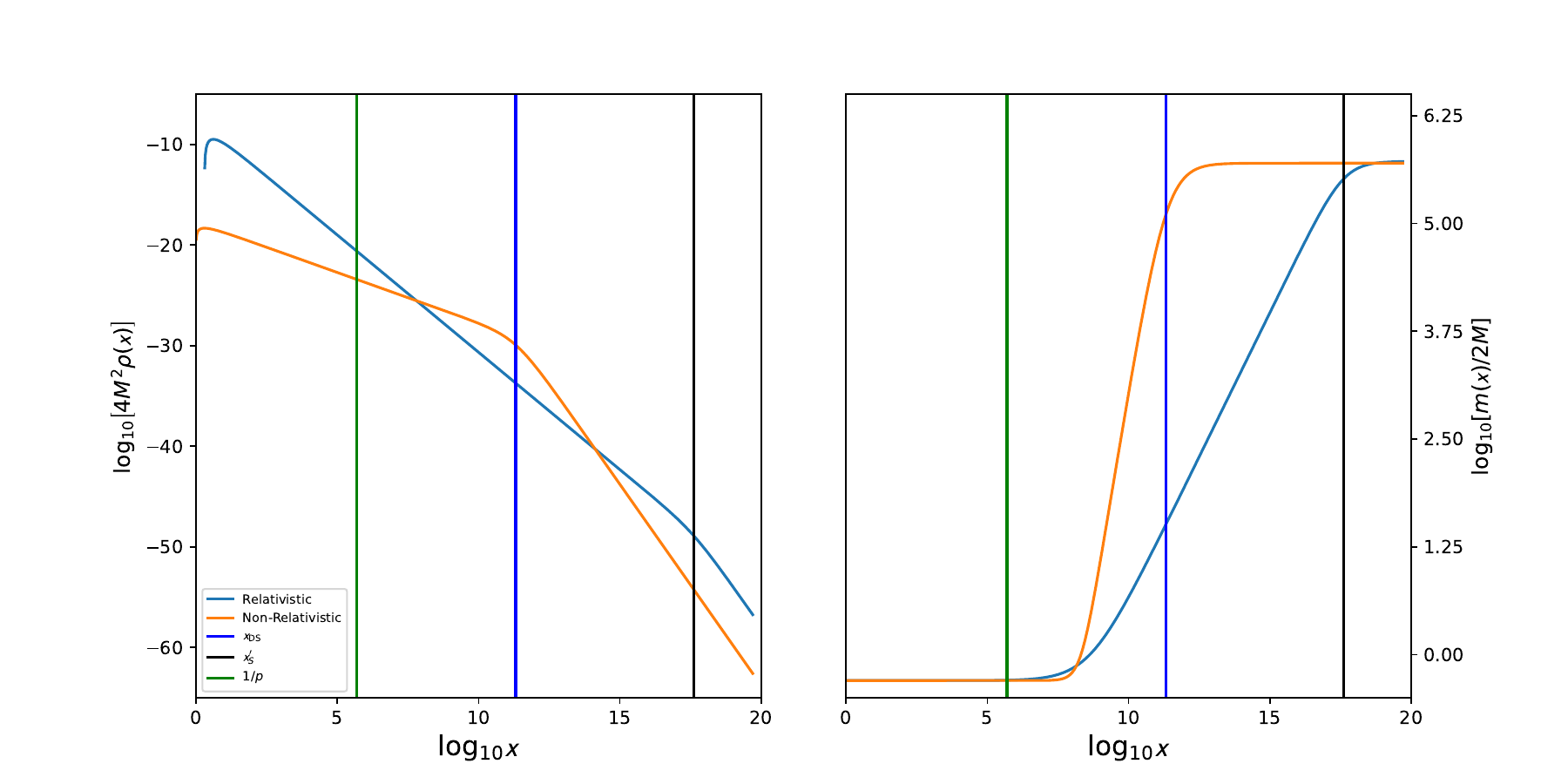}
    \caption{Radial profiles of the DM density $\rho(x)$ (left) and mass $m(x)$ for the non-relativistic and the relativistic DM formulations as a function of $x=r/2M$. The non-relativistic DM is taken from \cite{Cardoso:2021wlq} while the relativistic one is from \cite{Chakraborty:2024gcr}. Also shown as vertical lines are three dimensionless parameters $1/p, x_{\rm DS}$ and $x_S^\prime$ which are connected to three separate length-scales.}
    \label{fig:geom}
\end{figure} 
We have shown the constructions of the radial profiles of mass and density in~\ref{fig:geom}. It can be seen that the effect of the adiabatic contraction makes the mass function $m(x)$ grow less strongly for the relativistic profile compared to its non-relativistic counterpart. In the end the lengthscale $x_S^\prime$ turns out to be more accurate (compared to $x_{\rm DS}$) as a measure of the size of the DM halo for the relativistic profile, because $m(x)$ is seen to asymptote to $M + M_{\rm DM}$ only for $x> x_S^\prime$.

\section{Time Independent Perturbations: Love Number}\label{Sec:TIP_LN}

The review in~\ref{Sec:Geo_ECO} analyses the equilibrium background structure of the ECO-DM system. Let us now focus on perturbations to this system, starting with perturbations that are time-independent. The Love number is a well-known response parameter to time-independent perturbations for extended bodies when placed under an external inhomogeneous gravitational field. It has also been conclusively shown to be the leading-order tidal effect in the GW-phasing for coalescing binaries. It is therefore of interest to calculate the Love numbers of ECOs inside DM halos, if we want to understand the corresponding deviations from point-particle inspiral. As can be seen from previous literature, the computation of the Love number involves the solution of ordinary differential equations in the radial variable $x$. Written schematically, the differential equations can be written as
\begin{equation}\label{eq:odegen}
    \frac{d^2X(\boldsymbol{\theta};x)}{dx^2} + \alpha(\boldsymbol{\theta};x)\frac{dX(\boldsymbol{\theta};x)}{dx} + \lambda(\boldsymbol{\theta};x)X(\boldsymbol{\theta};x)=0~.
\end{equation}
Here $X$ is a generalised master variable which refers to perturbations while the vector $\boldsymbol{\theta}$ is used to represent all dependencies of the perturbation $X$ other than $x$. As we see, \ref{eq:odegen} is a homogeneous second-order equation and hence must have two linearly independent solutions in principle. After finding the solutions, some appropriate boundary conditions need to be imposed on them. Then the Love number can be computed from the large $x$ asymptotic behaviour of our solutions. Before proceeding with the calculations, let us highlight a couple of important points. First, we note from \cite{Chakraborty:2024gcr} that differential equations like \ref{eq:odegen} represent both Axial and Polar perturbations (Eqs. 124 and 138 of \cite{Chakraborty:2024gcr}). At the same time, it is also seen that \ref{eq:odegen} becomes analytically intractable when the exact expressions of $\alpha(\boldsymbol{\theta};x)$s and $\lambda(\boldsymbol{\theta};x)$s are substituted. The $\alpha(\boldsymbol{\theta};x)$ and $\lambda(\boldsymbol{\theta};x)$ terms have been shown to be explicit functions of the background fluid quantities $\rho, P_r$ and $P_t$. To overcome this problem of intractability, we follow the well-established scheme in the literature. We separate the perturbation schematically represented by $s(\boldsymbol{\theta};r)$ in a perturbative sense again into a part without DM and another part characterising DM as
\begin{equation}\label{eq:ssolndecomp}
    X(\boldsymbol{\theta};r) = X^{(0)}(\boldsymbol{\theta};r) + k\times X^{(1)}(\boldsymbol{\theta};r)~.
\end{equation}
Our perturbative expansion is in terms of a smallness parameter $k$, which was chosen to be $(M_{\rm DM}/r_S)$ in \cite{Chakraborty:2024gcr} which for a realistic astrophysical setting attains a value $p\sim 10^{-4}$. However, for our case we will keep $k$ an arbitrarily small parameter for the time being. The hierarchy of parameters by their absolute magnitude is also very important when talking about the asymptotic behaviour of any kind. We are interested in asymptotic behaviour attained within the halo, because the decomposition of the overall perturbation according to \ref{eq:ssolndecomp} is only sensible within the halo and not outside it. However, this poses an obvious question: which length scale is appropriate in order to consider asymptotic expansions at infinity? The answer is provided in~\ref{fig:geom} where a clear hierarchy of length-scales has been established, namely $1/p << x_{\rm DS} << x_S^\prime$. It is also readily seen that the appropriate regime to consider an $x\rightarrow\infty$ (or outer) asymptote is when $x\rightarrow x_B >>x_S^\prime$. Here, $x_B$ is just an arbitrary cutoff in $x$ to indicate the boundary of the DM halo. Physically, this means that for an astrophysical DM halo setting, the scale length $x_B$ can be treated as infinitely large compared to the ECO, and hence it makes sense to place an asymptotic observer at such a distance from the ECO. So when we talk of Love numbers, they are with respect to such asymptotic observers who themselves sit inside the halo. Let us now move to the solutions of \ref{eq:odegen}. If we substitute \ref{eq:ssolndecomp} into \ref{eq:odegen}, then we get two separate ordinary differential equations at the `0'th and first order in $k$, given respectively by 
\begin{eqnarray}
    X^{(0)''} + \alpha(\boldsymbol{\theta};x)X^{(0)'} + \lambda(\boldsymbol{\theta};x)X^{(0)}&=& 0\label{eq:odedecomp} \quad\text{and}\\
    X^{(1)''} + \alpha(\boldsymbol{\theta};x)X^{(1)'} + \lambda(\boldsymbol{\theta};x)X^{(1)}&=&\gamma(\boldsymbol{\theta};x)X^{(0)'}+  \beta(\boldsymbol{\theta};x)X^{(0)}~,\label{eq:odedecomp1}
\end{eqnarray}
where we have suppressed showing the independent variables of $X$ for brevity. Then, given that (\ref{eq:odedecomp}, \ref{eq:odedecomp1}), solving second-order differential equations, we can apriori write the general solution of the homogeneous parts of the differential equations involved in the following schematic way
\begin{equation}\label{eq:solodehmg}
    X^{(l)}(\boldsymbol{\theta};x) = A^l(\boldsymbol{\theta})y_1(\boldsymbol{\theta};x) +  B^l(\boldsymbol{\theta})y_2(\boldsymbol{\theta};x)~.
\end{equation}
Here the index $l$ represents the `$l$'th order (`0'th or first) of perturbations in $k$.  $A^k$ and $B^k$ are two arbitrary constants multiplying $y_1$ and $y_2$ respectively, which are linearly independent solutions.

\subsection{Solutions at `0'th order}\label{ssec:0thsoln}
We focus on the `0'th order equation given by \ref{eq:odedecomp} at first. Its solution is schematically given by
\begin{equation}\label{eq:solX0AB}
    X^{(0)}(\boldsymbol{\theta};x) = A^0(\boldsymbol{\theta})y_1(\boldsymbol{\theta};x) +  B^0(\boldsymbol{\theta})y_2(\boldsymbol{\theta};x)~.
\end{equation}
So far \ref{eq:solX0AB} is arbitrary in its nature, in that it does not specifically represent the metric functions of an ECO. To represent the perturbations for an ECO, $A^0$ and $B^0$ need to satisfy the ECO boundary conditions at the surface. For horizon reflectivity, we must have at the ECO surface given by $x_\epsilon =  1 + \epsilon$ the condition that $X^{(0)}\sim (1 + \mathcal{R})$. Given a reflectivity $\mathcal{R}$ we can now use the surface condition to relate $A^0(\boldsymbol{\theta})$ and $B^0(\boldsymbol{\theta})$. For our specific case we consider $\mathcal{R}= -1$ leading to
\begin{align}\label{A0B0reln}
B^0&=-A^0\left[\frac{y_1(\boldsymbol{\theta};\epsilon)}{y_2(\boldsymbol{\theta};\epsilon)}\right] \nonumber \\
&= -\kappa(\boldsymbol{\theta},\epsilon)A^0~.
\end{align}
A word about the nature of $y_1$ and $y_2$ is in order at this point. In deriving \ref{A0B0reln}, we have assumed that both $y_1$ and $y_2$ are regular at the ECO's surface. We also know from literature (eg, \cite{Hinderer:2007mb}) that a general pattern is observed for finite-sized objects, namely that one solution among $y_1$ and $y_2$ would blow up at $x=1$. Without loss of generality, we assume $y_2$ to be this solution. For BHs, $x=1$ is the horizon; therefore, the regularity requirement means that $B^0=0$. This condition is ultimately responsible for BHs having vanishing Love numbers. For ECOs, $x=1$ is inaccessible ($x\geq1+\epsilon$) meaning that both $y_1$ and $y_2$ survive. Then the quantity $\kappa(\boldsymbol{\theta},\epsilon)$ is nothing more than a shorthand of the term in the square brackets in \ref{A0B0reln} at this point. Its significance will be understood when we evaluate it for different cases explicitly later in this section. From \ref{A0B0reln} we also realise that in place of using the full expressions for $y_1$ and $y_2$, we can alternatively use their asymptotic expansions as $x\rightarrow 1+\epsilon$. We will refer to this situation as the `inner' asymptotic condition. We will see that the imposition of the inner asymptote form simplifies the computation of $\kappa$. Then substituting \ref{A0B0reln} into \ref{eq:solX0AB} we obtain
\begin{equation}\label{eq:solX0A}
    X^{(0)}(\boldsymbol{\theta};x) = A^0(\boldsymbol{\theta}) \left[y_1(\boldsymbol{\theta};x) - \kappa^A(\boldsymbol{\theta},\epsilon)  y_2(\boldsymbol{\theta};x)\right]~.
\end{equation}
We now come to the general method for the computation of the Love number, which involves the asymptotic form of \ref{eq:solX0A} when $x\rightarrow\infty$.  We will refer to this situation as the `outer' asymptotic condition. Physically, the Love number is a response to an extended body when placed in an external non-uniform gravitational field. This situation is not unique to gravitational fields, and the scenario has analogues in (for example) electromagnetism, where a finite-sized body is placed in an external electric or magnetic field. We can thus expect for our case that the outer asymptotic expansion of \ref{eq:solX0A} to have two contributions. The first is a part growing with $x$, which signifies the external potential. The second will be a decaying part, which signifies the potential produced by the object as a response. As is well-known from literature \cite{Chakraborty:2024gcr}, we have slightly different forms of $\alpha(\boldsymbol{\theta},x)$ and $\lambda(\boldsymbol{\theta},x)$ in \ref{eq:odegen} depending on the parity of the perturbations. Thus, for external gravitational fields, the solutions $y_1$ and $y_2$ have parity-dependent expressions. We highlight the case-specific details of the solutions in \ref{ssec:axial} and \ref{ssec:polar}. 

\subsection{Solutions at first order}\label{ssec:1stsoln}
Let us now move to highlight the generalised scheme for solving the first-order equation given by \ref{eq:odedecomp1}. This is an inhomogeneous equation and its general solution will be a linear combination of the homogeneous solutions and a particular solution. The homogeneous solution will be exactly given by the form of \ref{eq:solodehmg} with $l=1$. Then by substituting \ref{eq:solX0A} into \ref{eq:odedecomp1} we see that the coefficient $A^0(\boldsymbol{\theta})$ just scales out on the right-hand side. This means that the particular solution will also scale with $A^0$. Then the total solution is schematically given by
\begin{equation}\label{eq:solX1AB}
    X^{(1)}(\boldsymbol{\theta};x) = A^1(\boldsymbol{\theta})y_1(\boldsymbol{\theta};x) +  B^1(\boldsymbol{\theta})y_2(\boldsymbol{\theta};x) + A^0(\boldsymbol{\theta})\phi(\boldsymbol{\theta},x)~.
\end{equation}
where $\phi(\boldsymbol{\theta},x)$ is the particular solution in question. To compute $\phi(\boldsymbol{\theta},x)$ we employ the method to find inhomogeneous solutions to second-order differential equations called `variation of parameters' to \ref{eq:odedecomp1}. Assuming that $y_1$ and $y_2$ are a fundamental set of solutions to \ref{eq:odedecomp} (which can be checked by computing their Wronskian), a particular solution satisfying \ref{eq:odedecomp1} is given by
\begin{align}\label{eq:partsolnX1A}
    \phi(\boldsymbol{\theta};x) = -y_1(\boldsymbol{\theta};x) &\int dx \frac{\mathcal{S}(\boldsymbol{\theta};x )}{W(\boldsymbol{\theta,x})} y_2(\boldsymbol{\theta};x) \nonumber \\
    &+ y_2(\boldsymbol{\theta};x) \int dx \frac{\mathcal{S}(\boldsymbol{\theta};x )}{W(\boldsymbol{\theta,x})} y_1(\boldsymbol{\theta};x)~,
\end{align}
where $\mathcal{S}(\boldsymbol{\theta};x )$ is just the source term given by the right-hand side of \ref{eq:odedecomp1} and $W(\boldsymbol{\theta,x})$ is the Wronskian between the functions $y_1$ and $y_2$. Once again, the inner boundary condition at $x=1+\epsilon$ will help to establish a relation between the coefficients of \ref{eq:solX1AB}. The nature of the relation between $A^1$, $B^1$, and $A^0$ depends upon the explicit form of the particular solution $\phi(\boldsymbol{\theta},x)$. More specifically, the truncation of the relativistic density profile given by \ref{eq:rho_SchHeq} means from \ref{eq:partsolnX1A} that the solution $\phi(\boldsymbol{\theta},x)$ will go to $0$ at $x=2$. This feature sets the non-relativistic case where $\rho(x)$ is defined up to the ECO surface $x = 1+\epsilon$ on a different footing with respect to the relativistic one. Consequently, the imposition of the reflecting condition at the inner boundary means different things depending on whether the density is relativistic or non-relativistic. From \ref{eq:partsolnX1A} we obtain
\begin{align}\label{A1B1A0reln_NRel}
B^1(\boldsymbol{\theta}) &= -\kappa(\boldsymbol{\theta}, \epsilon) A^1 - \mu(\boldsymbol{\theta},\epsilon) A^0, \quad\text{where} \nonumber \\
\mu(\boldsymbol{\theta},\epsilon) &= \left[\frac{\phi(\boldsymbol{\theta},\epsilon)}{y_2(\boldsymbol{\theta},\epsilon)}\right]~.
\end{align}
In deriving \ref{A1B1A0reln_NRel} we made use of \ref{A0B0reln} to substitute for $\kappa(\boldsymbol{\theta},\epsilon)$. The absence of $\phi(\boldsymbol{\theta},x)$ at the ECO surface for the relativistic case implies a relation exactly like \ref{A0B0reln} in between $A^1$ and $B^1$. It must also be remembered that for the relativistic case, $A^1$ and $B^1$ are not the coefficients that will describe the general solution for $x\geq2$. This fact will have non-trivial consequences in \ref{ssec:axial} and \ref{ssec:polar}.

\paragraph{} This is as far as we can go with a generalised schematic approach. We will investigate case-specific scenarios and compute the associated Love numbers in \ref{ssec:axial} and \ref{ssec:polar}. In both these instances, our starting point will be given by the case-specific functions of \ref{eq:odedecomp} and \ref{eq:odedecomp1}.  We close the section with an illustration of asymptotes of background quantities, namely $m(x), \rho(x)$, and $p_t(x)$. We start with the outer region ($x_B>>x_S^\prime$) where we expect
\begin{align}\label{eq:rhomp_outasy}
    m(x\rightarrow x_B) &= (M + M_{\rm DM}) = M\left(1 + \frac{2}{p}\right) \nonumber \\
    \rho(x\rightarrow x_B) &= \frac{1}{2\pi}\left(\frac{\rm B}{x}\right)^4 \nonumber \\
    \frac{2p_t(x)}{\rho(x)} &= \frac{1}{x}\left(1 + \frac{2}{p}\right)~,
\end{align}
where we also assume $Mx>>m(x)$. This is a valid assumption as $x_B>> M_{\rm DM}$. $\rm B$ is a dimensionless constant that assumes different scaling for the relativistic and the non-relativistic case. The scalings are given by
\begin{align}\label{eq:Bout_asy}
    {\rm B}^4_{\rm rel} &= 2\pi (4M^2\rho_0) \times 2^{-q} {x_S^\prime}^{4-q} \nonumber \\
    {\rm B}^4_{\rm nrl} &= \frac{1+ x_{\rm DS}}{p}~.
\end{align}
Analogous expressions can also be worked out for inner asymptotes. We remember that the inner asymptote means $x\rightarrow x_i$ where $x_i=2 $ for the relativistic case and $x_i = 1$ for the non-relativistic case.
\begin{align}\label{eq:rhomp_inasy}
    m(x\rightarrow x_i) &= 1 \\
    \rho_{\rm R}(x\rightarrow x_i) &= 2^{-2q-w} (x-2)^w (4M^2\rho_0) \nonumber \\
    \rho_{\rm N}(x\rightarrow x_i) &= \frac{1}{2\pi p}\frac{x-1}{(x_{\rm DS} + 1)^2} \nonumber \\
    \frac{2p_t(x)}{\rho(x)} &= \frac{2x-1}{2(x-1)}~.
\end{align}
All expressions of the relevant functions have been reproduced from \cite{Chakraborty:2024gcr}.


\subsection{Axial Perturbations}\label{ssec:axial}
In this section, we focus on the odd parity perturbations, starting with the `0'th order. The exact solutions to \ref{eq:odedecomp} was given in \ref{ssec:0thsoln} by \ref{eq:solX0A} where
\begin{align}\label{eq:y1y2axial}
    y_1(\boldsymbol{\theta};x) &= x^2 (x-1) \nonumber \\
    y_2(\boldsymbol{\theta};x) &=\bigg(\frac{1+2 x+6 x^2-12 x^3}{3x}\bigg)-4 x^{2}(x-1)\log(1-\frac{1}{x})~. \nonumber \\
\end{align}
In \ref{eq:y1y2axial}, the vector $\boldsymbol{\theta}$ represents $\ell=2$-Axial solutions. We can now take an outer asymptotic expansion of \ref{eq:y1y2axial} and substitute the forms into \ref{eq:solX0A}. It is then clearly seen that the whole solution has a growing part behaving as $x^3$ and a decaying part going as $1/x^2$ at leading order. 
\begin{equation}\label{eq:solaxial0}
    h_0^{(0),a}(\ell=2;x) = A^{0,a} \left[x^3 - \kappa^a(\epsilon) \left(\frac{-1}{5x^2}\right) \right]~.
\end{equation}
The ratio of the growing and the decaying coefficients gives the Love number. Hence, for Axial perturbations without DM, the Love number is given by
\begin{align}\label{eq:k2a0}
&k_2^{a0} = \left(\frac{1}{5}\right)\kappa^a(\epsilon).
\end{align}
In both \ref{eq:solaxial0} and \ref{eq:k2a0}, the superscript `a' refers to axial perturbations. We can now see that the factor $\kappa^a(\epsilon)$ is just related to the $\ell=2$ Love number by a constant factor. We also note that, unlike \cite{Chakraborty:2024gcr}, \ref{eq:k2a0} is dimensionless and hence does not have an $M^5$ scaling. To compute $\kappa^a(\epsilon)$ we just use the inner asymptotic form of $y_1$ and $y_2$ from \ref{eq:y1y2axial}, leading to 
\begin{equation}\label{eq:kappa_a}
    \kappa^{a}(\epsilon) = -\epsilon + 4\left(\frac{13}{12} + {\rm log}\epsilon\right)\epsilon^2 + \mathcal{O}(\epsilon^2)~.
\end{equation}
Let's focus on the particular solution $\phi(\ell=2,x)$ for Axial perturbations. Without loss of generality, we can assume it has an outer asymptotic form given by 
\begin{equation}\label{eq:phiexp}
    \phi(\ell =2, x) = \sum_i \alpha_i x^i~.
\end{equation}
The growing and decaying parts of $\phi$ are therefore represented by those indices $i$ which are positive and negative, respectively. To compute the Love number of the composite object (ECO in DM halo), we have to separate out the $x^3$ and $1/x^2$ parts of this perturbation. In other words, we are interested in those parts of the field which mimic compact object geometry. The overall perturbation will then acquire an asymptotic form given by
\begin{align}\label{eq:solaxial1}
    &h_T^{a}(\ell=2;x) = A^{0,a}x^3 + B^{0,a} \left(\frac{-1}{5x^2}\right)  \nonumber\\
    +\quad&k\left[ A^{1,a}_\rho x^3 + B^{1,a}_\rho \left(\frac{-1}{5x^2}\right) + A^{0,a} \left(\alpha_3^a x^3 + \alpha_{-2}^a x^{-2}\right)\right]~.
\end{align}
The coefficients $A^1_\rho, B^1_\rho$ proportional to $p$ have been denoted by a subscript $\rho$ which indicates a difference in values depending on whether $\rho$ is relativistic or non-relativistic. Fortunately, a unified representation of the cases is possible and is given as follows. If $\phi$ does not extend to the ECO surface, then $A^1$, $B^1$ no longer represent the constants beyond $x=2$. Let $A_\rho^1$ and $B_\rho^1$ be the generalised constants. Then the requirement of continuity and differentiability across $x=2$ means
\begin{gather}\label{cond-contdiff}
 \begin{pmatrix} A^1_\rho  \\ B^1_\rho  \end{pmatrix}
 =
 \begin{pmatrix} A^1  \\ B^1  \end{pmatrix}
 -A^0
  {\begin{pmatrix}
   y_1 &
   y_2 \\
   y_1^\prime &
   y_2^\prime 
   \end{pmatrix}}^{-1}_{x=2}
   {\begin{pmatrix} \phi  \\ \phi^\prime  \end{pmatrix}}_{x=2}~.
\end{gather}
\ref{cond-contdiff} is then the expression of $A_\rho^1$ and $B_\rho^1$ in the relativistic case when $\phi$ terminates at $x=2$. For the non-relativistic case, $A_\rho^1 = A^1$ and $B_\rho^1=B^1$.
Writing out the components from \ref{cond-contdiff}, we can pack all information into a single equation given by
\begin{align}\label{eq:abrho}
    A_1^\rho &=  A_1 - \Theta(A^0\chi_1) \nonumber \\
    B_1^\rho &=  B_1 - \Theta(A^0\chi_2)~,
\end{align}
where $\chi_i$s are constant factors of $y_1,y_2$ and their derivatives at $x=2$ and $\Theta$ is a parameter which is $1$ for relativistic case and $0$ for non-relativistic case. Then from \ref{eq:solaxial1}, the Axial Love number is formally calculated by the ratio of the $1/x^2$ and $x^3$ coefficients given by
\begin{equation}\label{eq:kappa_a-gen}
    k_2^a = \left[\frac{-B^{0,a}/5 + k(-B^{1,a}_\rho/5 + A^{0,a}\alpha_{-2}^a)}{A^{0,a} + k(A^{1,a}_\rho + A^{0,a}\alpha_{3}^a)}\right] ~.
\end{equation}
For a given DM density profile, the $A^0, B^0$ coefficients are related among themselves by \ref{A0B0reln} while the $A^0, B^0$ coefficients are related through \ref{A1B1A0reln_NRel}. Then in \ref{eq:kappa_a-gen} we substitute for $A^1_\rho,B^1_\rho$ using \ref{eq:abrho}. Combined with \ref{eq:kappa_a} gives us the following expression to leading order in $p$
\begin{align}\label{eq:k2a}
&k_2^{a} = k_2^{a0} + \frac{k}{5}\left[\mu + \Theta(\chi_1 + \kappa\chi_2) + 5\alpha_{-2}^a -\alpha_3^a\kappa\right].
\end{align}
The functional dependencies on $(\boldsymbol{\theta},\epsilon)$ have been suppressed on the right-hand side of \ref{eq:k2a} for brevity. Our result for the Love number of an ECO in DM is given by \ref{eq:k2a}, which valid result irrespective of whether the density is relativistic or non-relativistic. Looking at \ref{eq:k2a}, we see that the overall contribution (to the Love number) has a DM independent part and a DM dependent part proportional to $p$. Given our decomposition of perturbations following \ref{eq:odedecomp} and \ref{eq:odedecomp1}, this is quite expected. Additionally, the DM dependent portion also depends upon the functions $\kappa$ defined on the ECO surface. An interesting feature of \ref{eq:k2a} is the transition from relativistic to non-relativistic regime and vice-versa. For the relativistic case $\mu=0,$ $\Theta\neq0$ while the opposite is true for the non-relativistic case.

\subsection{Polar Perturbations}\label{ssec:polar}
Let us now turn to the even parity perturbations at `0'th order. The forms of $y_1$ and $y_2$ are now given for $\ell=2$ by
\begin{align}\label{eq:y1y2axial1}
    y_1(\boldsymbol{\theta};x) &=  12x^2 \left(1 - \frac{1}{x}\right) \nonumber \\
    y_2(\boldsymbol{\theta};x) &=4x^2 \left(1 - \frac{1}{x}\right) \left[ \frac{(2x-1)(1+6x-6x^2)}{8x^2(x-1)^2} - \frac{3}{2}{\rm log}\left(1 - \frac{1}{x}\right) \right]~. \nonumber \\
\end{align}
We recognise that these are just the Associated Legendre functions of the first and the second kind, namely $P_2^2(2x-1)$ and $Q_2^2(2x-1)$. We note here an apparent dissimilarity with the axial case. Whereas in the axial case \ref{eq:odedecomp} has known solutions in terms of special functions only for the $\ell = 2$ case, in the polar case the Associated Legendre functions are valid solutions for arbitrary $\ell\geq 2$. Then an equation analogous to the Axial case in \ref{eq:solaxial0} is given by
\begin{equation}\label{eq:solpolar0}
    h_0^{(0),p}(\ell=2;x) = A^{0,p} \left[12x^2 - \kappa^p(\epsilon) \left(\frac{1}{5x^3}\right) \right]~.
\end{equation}
We see that for Polar perturbations, the order of the growing and decaying parts is reversed. The growing part behaves as $x^2$ while the decaying part goes as $1/x^3$. Then the ratio of the growing and the decaying coefficients is
\begin{align}\label{eq:k2p0}
&k_2^{p0} = -\left(\frac{1}{60}\right)\kappa^p(\epsilon).
\end{align}
Then the value of $\kappa^p(\epsilon)$ is given analogously by
\begin{equation}\label{eq:kappa_p1}
    \kappa^{a}(\epsilon) = 24\epsilon^2 +  \mathcal{O}(\epsilon^3)~.
\end{equation}
The subsequent treatment for the polar case is exactly the same as for the axial case given by \ref{eq:phiexp} and \ref{eq:solaxial1}, with the notable difference that the Love number is now defined as the ratio between the $1/x^3$ and $x^2$ coefficients. Analogous to \ref{eq:kappa_a} we then obtain
\begin{equation}\label{eq:kappa_p}
    k_2^p = \left[\frac{B^{0,p}/5 + k(B^{1,p}_\rho/5 + A^{0,p}\alpha_{-3}^p)}{12A^{0,p} + k(12A^{1,p}_\rho + A^{0,p}\alpha_{2}^p)}\right] ~.
\end{equation}
Then we perform a similar calculation to \ref{eq:k2a}, wherein we combine \ref{eq:kappa_p} with \ref{eq:abrho} and \ref{eq:k2p0} to derive the Love number for the Polar case at linear order in $p$
\begin{align}\label{eq:k2p}
&k_2^{p} = k_2^{p0} - \frac{k}{60}\left[\mu +\Theta(\chi_1 + \kappa\chi_2)- 5\alpha_{-3}^p -\frac{1}{12}\alpha_2^p\kappa\right].
\end{align}
It is readily seen that \ref{eq:k2p} has exactly the same structure as \ref{eq:k2a}, except for values of the coefficients.

\subsection{Evaluating the Love numbers}\label{ssec:eva_love}

Our only remaining task now is to explicitly evaluate for the functions $\mu$ and the relevant $\alpha$s, which can be done by calculating the inner and outer asymptotic form of the particular solution $\phi(\boldsymbol{\theta},x)$ from \ref{eq:partsolnX1A}. In order to make use of \ref{eq:partsolnX1A}, we need to simplify its right-hand side by substituting the appropriate expansions of the corresponding terms given by \ref{eq:ax-terms} in \ref{ssec:sta}. We substitute the outer expansions given by \ref{eq:rhomp_outasy} to compute $\alpha$s and the inner expansion given by \ref{eq:rhomp_inasy} to compute $\mu$ and $\chi$s in \ref{eq:k2a} and \ref{eq:k2p}. In this way, while we do not have an exact expression of $\phi(\boldsymbol{\theta},x)$, we can work around the problem with its outer and inner asymptotic forms.  Let us focus on the evaluation of the $\alpha$s. They are namely $\alpha_2^p,$ $\alpha_3^a$ representing the growing parts and $\alpha_{-3}^p,$ $\alpha_{-2}^a$ representing the decaying parts in the particular solution. We find that both $\alpha_2^p$ and $\alpha_3^a$ vanish. Physically, this means that any asymptotically growing part of the perturbation must be from an external field. A finite-sized ECO-DM system can only generate an asymptotically decaying response. This behaviour should be expected. Meanwhile, the values for $\alpha_{-3}^p$ and $\alpha_{-2}^a$ are given by
\begin{align}\label{eq:alphas}
    \alpha_{-3}^p &= \frac{1}{k}\left[\frac{28}{10}\left(\frac{1}{p}\right)^2 - \frac{3}{2}\left(\frac{1}{p}\right) - \frac{7}{5}{\rm B}^4p\right]~, \nonumber \\
    \alpha_{-2}^a &= \frac{1}{k}\left[\frac{2}{15}\left(\frac{1}{p}\right) + \frac{{\rm B}^4}{p}\left(\frac{63-60 {\rm log}(x)}{75}\right)\right]~.\nonumber \\
\end{align}
The values in \ref{eq:alphas} raise a crucial point of the paper. Looking at the dependencies, we see that the values have scalings of $1/p$ and $1/p^2$, all of which apparently make the perturbative treatment self-inconsistent. This inconsistency stems from a subtle point discussed in Sec. 5.1.1 of \cite{Chakraborty:2024gcr}. It turns out that the perturbative equation at first order, namely \ref{eq:odedecomp1}, has a second arbitrary choice associated with it. This choice is of an effective length-scale which is introduced as the variable $R$ in Eq.126 of \cite{Chakraborty:2024gcr}. For our case, since we use the reduced variable $x = r/2M$ for \ref{eq:odedecomp1} it automatically sets $R=2M$, which then makes the perturbative treatment self-inconsistent. To make sense of the perturbation, we need to choose a length scale appropriate to the choice of $k$. In \cite{Chakraborty:2024gcr}, this length scale is chosen as $R=r_s$. However, for our purpose let us also keep this choice arbitrary such that $2M/R = f$. Consequently, the terms in \ref{eq:k2a} and \ref{eq:k2p} which originate from the particular solution $\phi$ need to be rescaled by appropriate factors of $f$. Thus $ \alpha_{-3}^p$ will have to be multiplied by a factor of $f^3$ while the same factor for $\alpha_{-2}^a$ is $f^2$. The scaling applies to one more coefficient, namely to ${\rm B}^4$. This is because we use the factor of $M^2$ to make $\rho$ dimensionless in \ref{eq:rhox_H} and \ref{eq:rhox_NR}. This means that the factor ${\rm B}^4$ should also receive an additional rescaling factor of $f^2$. Combining all the scalings, we are then left with
\begin{align}\label{eq:alphascale1}
    \alpha_{-3}^p &= \frac{f^3}{kp^2}\left[\frac{28}{10} - \frac{3}{2}p^2 - \frac{7}{5}\left({\rm B}^4f^2\right)p^3\right]~ \nonumber~, \\
    \alpha_{-2}^a &= \frac{f^2}{kp}\left[\frac{2}{15} + \left({\rm B}^4f^2\right)\left(\frac{63-60 {\rm log}(x)}{75}\right)\right]~.\nonumber \\
\end{align}
The choice $f=k$ means $\alpha_{-2}^a$ and $\alpha_{-3}^p$ give the same scaling for the axial and polar cases. Then an appropriate choice of $k$ requires the condition that the quantity given by ${\rm B^4}f^2$ respects the perturbative sense of the coefficients. From the expressions of ${\rm B}^4$ in \ref{eq:Bout_asy} it turns out that the choice $f = 1/{(x_{\rm DS}})^2$ satisfies the above condition for the relativistic case, while the same for the non-relativistic case is satisfied by $f = \sqrt{p/x_{\rm DS}} = 2/\sqrt{x_S^\prime}$. All that remains now is the evaluation of the coefficients $\chi_1,\chi_2$ and $\mu$. It turns out that both $\phi$ and its derivative are $0$ at $x=2$ because of terms proportional to powers of $(x-2)$. The origin of these terms can be traced to the inner asymptote behaviour of $\rho$ in the relativistic case from \ref{eq:rhomp_inasy}. As a consequence $\chi_{1}$ and $\chi_{2}$ both vanish. The final quantity to evaluate is $\mu$, which, as we know, vanishes for the relativistic case. For the non-relativistic cases, the computation gives us different values for polar and axial cases as follows
\begin{align}\label{eq:mu}
    \mu^p &=\frac{1}{2}{\rm c}_\mu^2 \times   \kappa ^2 \epsilon  (1+8 \epsilon  \log \epsilon)~, \nonumber \\
    \mu^a &=2 {\rm c}_\mu \times  \kappa  \epsilon  (1-\log \epsilon )~, \quad\text{where} \nonumber \\
    {\rm c}_\mu &= \frac{1}{p(1+x_{\rm DS})^2}~.
\end{align}
The calculation of the Love numbers is completed by substituting \ref{eq:alphascale1} and \ref{eq:mu} into \ref{eq:k2a} and \ref{eq:k2p} for the axial and polar parts, respectively.

\section{Love numbers for truncated halos}\label{sec:TH_Love}
In the previous sections, we assumed that the asymptotic behaviour of the DM halo is attained when the mass of the ECO-halo system attains the value $M + M_{\rm DM}$. However, this may not always be the case, as the DM particles might not be gravitationally bound to the system beyond a cutoff length-scale $x_B$ such that $x_B< x_S^\prime$ or  $x_B< x_{\rm DS}$. In such situations, the total mass of the system is no longer given by $M + M_{\rm DM}$. Instead, the total mass of the system would be (depending upon the value of $x_B$) less than $M_{\rm DM}$, which would then just become another parameter in $\rho(x)$. Physically, this system represents a truncated halo. We are interested in the process to compute the Love Numbers of such truncated halos. We recognise that this scenario is exactly similar to those of compact objects like white dwarfs (WDs) or neutron stars (NSs). Accordingly we have a region interior to the halo and a region exterior to it. Just like in the case of NS, the perturbation in the interior will satisfy \ref{eq:odegen}, while that in the exterior region will be like \ref{eq:odedecomp}. Let $\Bar{X}_{\rm in}$ and $\Bar{X}_{\rm ex}$ denote the perturbations in the interior and exterior respectively. Then given that $\Bar{X}_{\rm ex}$ obeys \ref{eq:odedecomp}, we can express
\begin{equation}\label{eq:TH-ext_soln}
    \Bar{X}_{\rm ex} = {\rm P}y_1(x^\prime) + {\rm Q} y_2(x^\prime)~,
\end{equation}
where the primed variables in the arguments indicate that for \ref{eq:TH-ext_soln} the radial variable $r$ has to be scaled to $x^\prime$ by the corresponding mass of the truncated halo $x^\prime = r/M^t_h$. The truncation point is $x^\prime_B = R_h^t/M_h^t$, where $R_h^t$ is the radius of the incomplete halo. Once again, the condition of continuity and differentiability has to be enforced across the halo boundary $x^\prime_B$, which gives us a relation just like \ref{cond-contdiff} 
\begin{gather}\label{cond-contdiff2}
 \begin{pmatrix} {\rm P}  \\ {\rm Q}  \end{pmatrix}
 =
  {\begin{pmatrix}
   y_1 &
   y_2 \\
   y_1^\prime &
   y_2^\prime 
   \end{pmatrix}}^{-1}_{x=x^\prime_B}
   {\begin{pmatrix} \Bar{X}_{\rm in}  \\ \Bar{X}_{\rm in} ^\prime \end{pmatrix}}_{x=x^\prime_B}~.
\end{gather}
\begin{figure}[h]
    \centering{
    \includegraphics[width=0.8\linewidth]{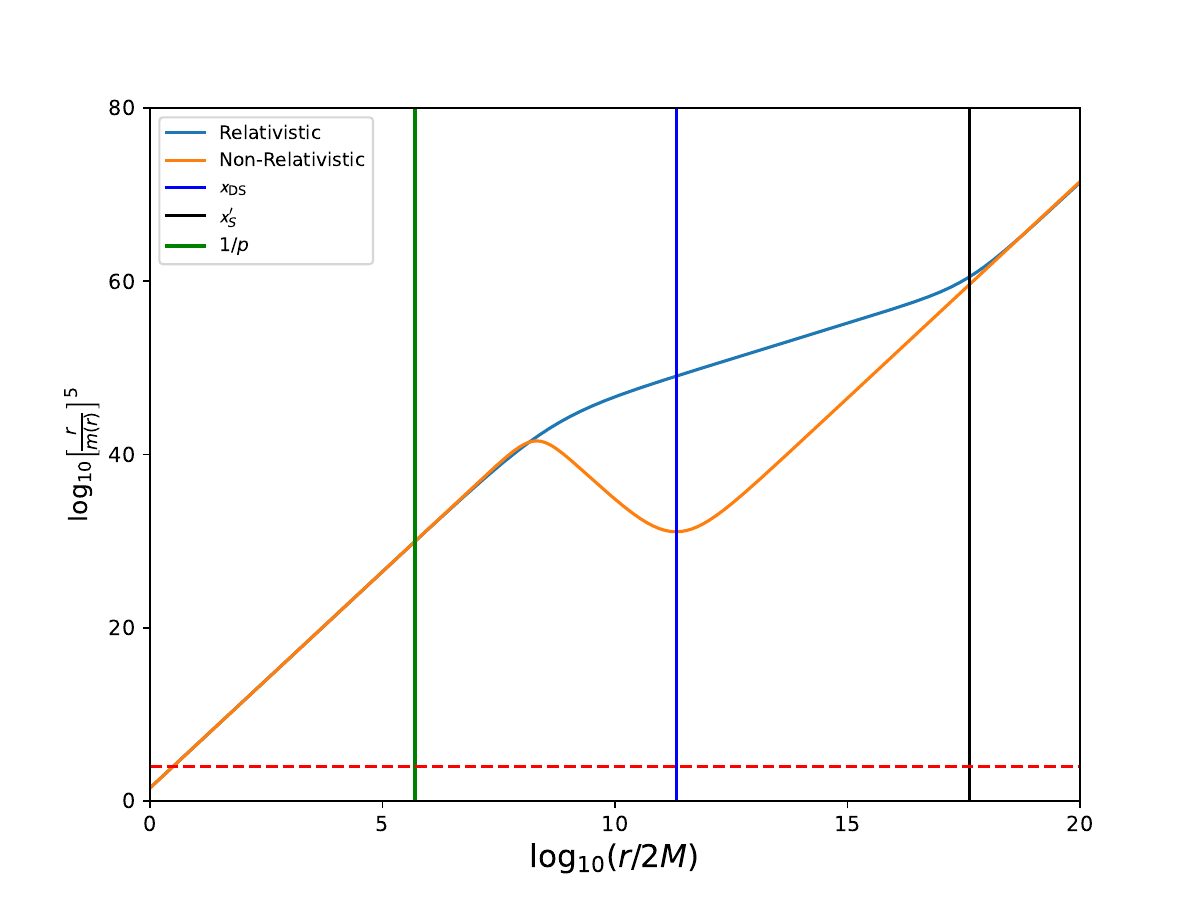}}
    \caption{This plot shows the scaling of the factor $(r/m(r))^5$ as a function of $r$, for the relativistic and non relativistic halos of~\ref{fig:geom}. This factor is ultimately responsible for determining the order of magnitude of the truncated halos. The dashed red line at the bottom shows the maximum possible order of magnitude value of Love numbers attained by astrophysical NSs.}
    \label{fig:LTH}
\end{figure}
The values of $\Bar{X}_{\rm in}$ and its derivative at $x=x^\prime_B$ fix ${\rm P}$ and ${\rm Q}$ uniquely. We can now easily compute the Love numbers by once again taking the ratio of the asymptotically growing and decaying parts of $\Bar{X}_{\rm ex}$ at leading order.  For both perturbations, this implies
\begin{align}\label{eq:k2th}
    k_{t2}^{a,p} &= \mathcal{F}_{a,p} \times \left(\frac{y_1}{y_2}\right)_{x^\prime_B} \times \left(\frac{(y_1^\prime/y_1) - (\Bar{X}_{\rm in} ^\prime/ \Bar{X}_{\rm in} )}{(y_2^\prime/y_2) - (\Bar{X}_{\rm in} ^\prime/ \Bar{X}_{\rm in} )}\right)_{x^\prime_B}~,
\end{align}
where $\mathcal{F}_{a,p}$ is a constant factor which assumes the value $1/5$ for axial perturbations and $1/60$ for polar perturbations. We have verified that the third term of \ref{eq:k2th} is almost a constant for a given halo with respect to its truncation point $x^\prime_B$. The remaining factor is proportional to the factor 
$(x^\prime_B)^5 = (R_h^t/M_h^t)^5$ for $x^\prime_B>>1$. This presents a one-to-one correspondence with NSs (see, for example, Eq. 13 of \cite{Rezzolla:2016nxn}) whose dimensionless Love number also scales with the factor $(R/M)^5$. This presents an interesting question. Given that truncated halos behave identically to NSs functionally, is it possible that one might be misidentified for the other? We realise that this is possible if NSs and the truncated halos have Love numbers close to each other. For NSs, depending upon their masses, this value is typically $\sim 10^3-10^4$. Accordingly, we vary the truncation point with $r$. Clearly for the halos we consider, the factor $\left(\frac{r}{m(r)}\right)^5$ will give us the order of magnitude of the Love numbers. We have plotted this factor as a function of the radial variable $r$ in~\ref{fig:LTH}. For choices of the halo boundary such that $x^\prime_B>>1$, we see that the order of magnitude of the Love numbers will be $10^{40}-10^{50}$ which is nowhere near those attained by NSs. It is also seen from~\ref{fig:LTH} that under the asymptotic condition given by $x^\prime_B>>1$, the Love numbers of the halo cannot become the same order as those of NSs. Therefore, the possibility of misidentification does not exist. Physically, this just means that the truncated halos can never be so compact as to mimic NSs, which is what should be expected.

\section{Time dependent Perturbation: Echo}\label{Sec:TDP_Echo}
Until now, we were only concerned with perturbations independent of time. In this section, we will consider time-dependent perturbations. For time-dependent perturbations in spherical symmetry, it is well-known that the perturbation equations can be reduced to the Regge-Wheeler form in terms of a master variable. Accordingly, the $(r,\phi)$ component of the perturbed Einstein equations is cast into Regge-Wheeler form with a source term. It turns out \cite{Chakraborty:2024gcr} that for the case of Axial perturbations the master variable can be defined as $\Psi = (\sqrt{fg}/r)h^a$, where $h^a$ is the Axial perturbation defined in ~\ref{ssec:axial}, but now is explicitly a function of time. Additionally, we also introduce the tortoise variable $r_*$ in the usual way such that $dr_* = dr/(\sqrt{fg})$ This yields the familiar radial equation for perturbations given by
\begin{figure}[h]
    \centering{
    \includegraphics[width=0.9\linewidth]{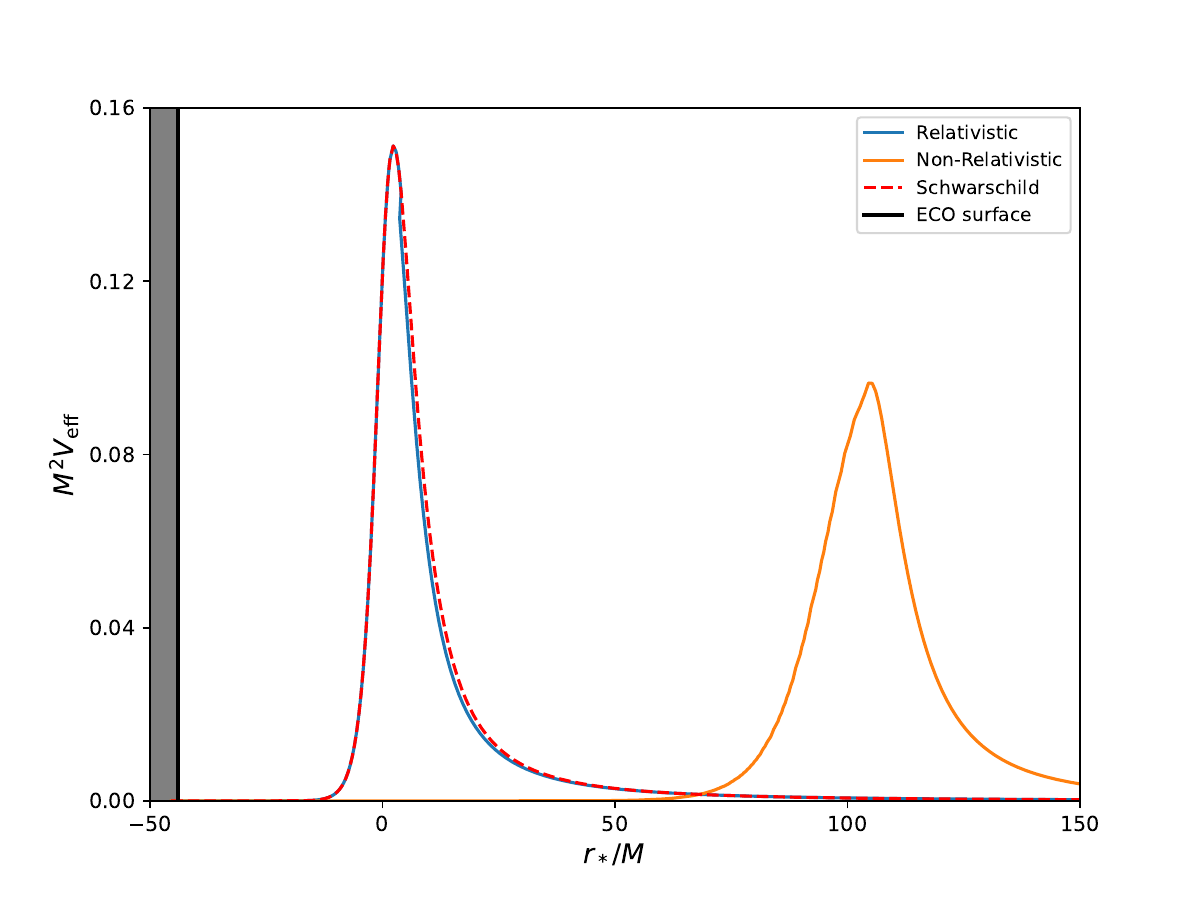}}
    \caption{Effective potential $V_\mathrm{eff}$ from our chosen ECO-DM with relativistic and non-relativistic DM profiles plotted as a function of the tortoise radial coordinate $r_*$. The Schwarschild barrier is shown for reference. The cut-off of the relativistic DM at $r/M=4$ appears as a small irregular feature. Note the displacement of the non-relativistic barrier. The gray shaded region on the far right is the inaccessible region below the ECO surface. See text for details.}
    \label{fig:Pots}
\end{figure} 
\begin{align}\label{eq:pert_master}
\frac{\partial^{2}{\Psi}}{\partial r_{*}^{2}} - \frac{\partial^{2}{\Psi}}{\partial t^2}- V_\mathrm{eff}(r){\Psi}
=\frac{4f\sqrt{fg}}{r}\left(\frac{\bar p_{\rm t}-\bar p_{\rm r}}{\bar \rho+\bar p_{\rm t}}\right)V^{\rm{up}}~.
\end{align}
From this point onward, it is actually easier to retain $t$ and $r_*$ as coordinates to obtain the echo signal. Unless explicitly stated, we will not use the coordinate $x$. In \ref{eq:pert_master} $V^{\rm{up}}$ is a source term which vanishes in the absence of perturbations to the fluid 4 velocity. This is the case we are interested in, namely ECO perturbations without any fluid perturbations in the DM. With these implementations, \ref{eq:pert_master} reduces to the following well-known Regge-Wheeler form given by 
\begin{equation}\label{eq:RW_master}
    \left(\partial_t^2 -\partial_{r_*}^2\right)\Psi(t,r_*) - V_\mathrm{eff}(r)\Psi(t,r_*) = 0,
\end{equation}
where $V_{\rm eff}(r)$ denotes the effective Regge-Wheeler potential with anisotropic stress contributions seen by the perturbation. It is given by
\begin{align}\label{eq:Veff_pert}
V_\mathrm{eff}(r)= f\left[\frac{\ell(\ell+1)}{r^{2}}- \frac{6m(r)}{r^3}   +4\pi\left(\rho + 4 P_t\right)\right]~,
\end{align}
where we have implemented vanishing radial pressure term $\Bar{p}_r$.
We have plotted the behaviour of  $V_\mathrm{eff}$ in~\ref{fig:Pots} as a function of the tortoise coordinate $r_*/M$ for the cases of the relativistic and non-relativistic DM profiles along with the Schwarzschild barrier. Our chosen system is one where $M = 20 M_\odot$, $M_\mathrm{DM} = 50 M_\odot$ and $r_S = 100 M_\odot$. Our choices have been purposely made in the hope of maximising the difference among the different kinds of barrier potentials. For the relativistic case,  $r\leq4M$ means $\rho(r) = 0$ and $m(r) = M$. So for that region we see from \ref{eq:Veff_pert} that $V_{\rm eff}$ reduces to the Schwarzschild potential. Additionally, the tortoise coordinates for the relativistic case also coincide with those of the Schwarzschild barrier in this region. This is seen from~\ref{fig:Pots} as a total overlap of the red and blue curves below $r = 4M$. Furthermore, it also turns out that $V_{\rm eff}$ is dominated by the centrifugal component of the potential, meaning that in the relativistic DM case, the light ring continues to be at $r = 3M$. At $r>4M$, both $V_{\rm eff}$ and $r_*$ deviate from their Schwarzschild counterparts. For the relativistic profile, this deviation is also almost imperceptible, which means that the relativistic DM profile produces a potential barrier that is virtually indistinguishable from that of its Schwarzschild counterpart. The situation changes, however, when we work with the non-relativistic DM profile. Unlike the relativistic $\rho(r)$, it extends all the way to the ECO's surface. It is known from previous work \cite{Cardoso:2021wlq} that for the non-relativistic DM profile, the light ring radius gets modified to 
\begin{equation}\label{eq:lr_nr}
    r_{\rm lr} = 3M \left(1 + \frac{MM_{\rm DM}}{r_S^2}\right)~.
\end{equation}
In our case it implies a shift of the light ring to the right by a factor $10\%$. Additionally, $r_*$ now gets contributions from the $r<4M$ region as well and therefore becomes completely different with respect to the Schwarzschild case. The combination of these two effects pushes the barrier potential in the non-relativistic case much further to the right in the $r_*$ coordinate, as is seen from~\ref{fig:Pots}. The appropriate Schwarschild limit is recovered if $r_S>>\sqrt{MM_{\rm DM}}$ as can be seen from \ref{eq:lr_nr}. 

\bigskip

\noindent Let us now return to the problem at hand, namely, to figure out the evolution of perturbations governed by \ref{eq:RW_master}. If we now enforce the ingoing boundary condition upon \ref{eq:RW_master}, we get the discrete spectrum of QNMs corresponding to the BH perturbation. However, for non-ingoing boundary \ref{eq:RW_master} cannot be solved analytically. To study the evolution of perturbations, one needs to perform an evolution of initial data on a Cauchy hypersurface. In order to proceed it is instructive to make a transformation into double-null coordinates $u\vcentcolon= (t - r_*)/M; v\vcentcolon= (t + r_*)/M$ following previous literature in this direction \cite{Chakraborty:2017qve}. With this transformation, the master equation ~\ref{eq:RW_master} becomes
\begin{equation}\label{eq:RW_uv}
    4\partial_u\partial_v\Psi(u,v) - M^2V_\mathrm{eff}(u,v)\Psi(u,v) = 0,
\end{equation}
\begin{figure}[h]
    \centering{
    \includegraphics[width=0.9\linewidth]{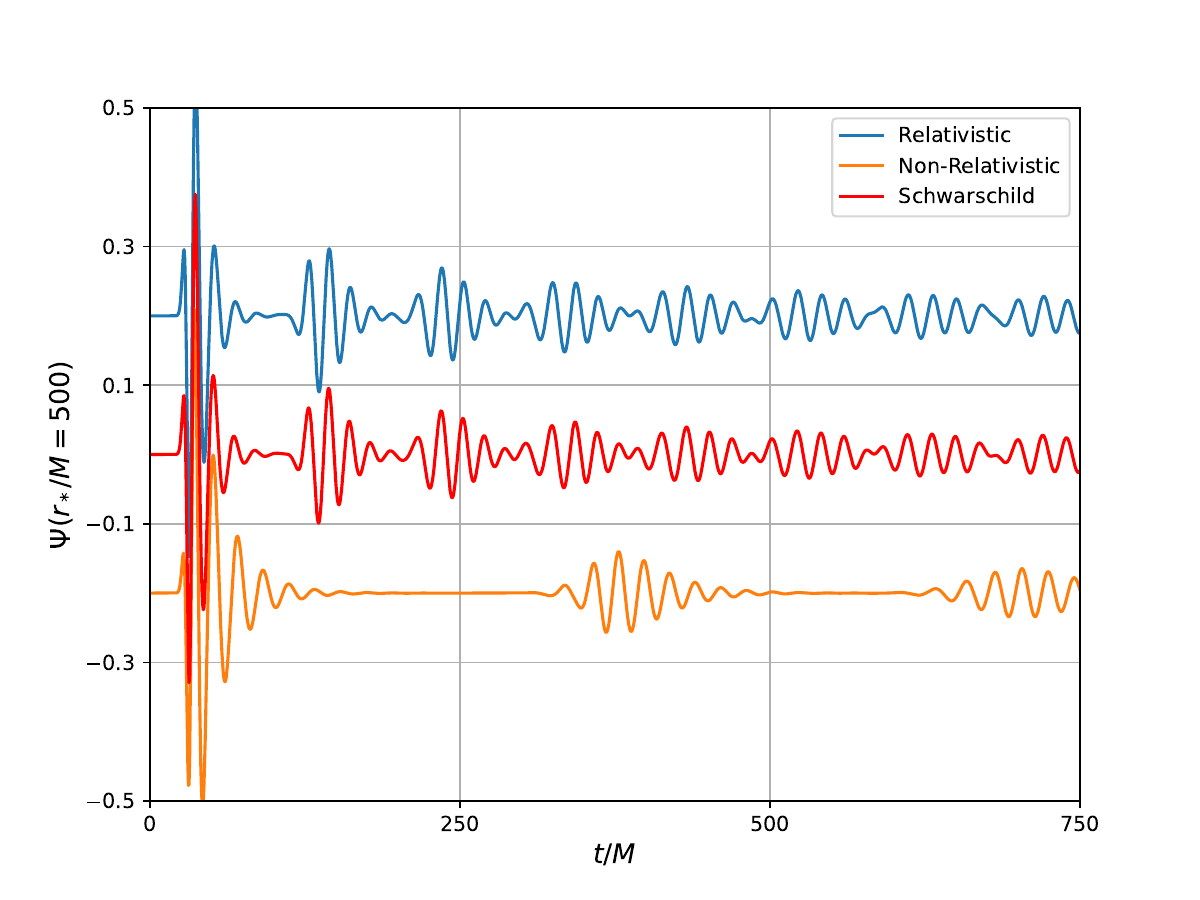}}
    \caption{This plot shows the GW-Echoes from our ECO-DM systems. The Schwarschild case represents the potential of an ECO without any DM. The observation point is taken to be at $r_*/M = 500 $. We stack the echo wave-trains on top of each other in order to have better visualisation.}
    \label{fig:Echo1}
\end{figure}
This equation has to be solved numerically. To do this, we first discretise the $u,v$  plane in steps of $h$ with $h<<1$. Then using ~\ref{eq:RW_uv} and the fact that ${\rm exp}[(Mh)\partial_t]$ is a time evolution operator in step of $Mh$, we can write the evolution of $\Psi(u,v)$ to leading order of $h$ as
\begin{align}\label{eq:Psi_evol}
    &\Psi(u,v) = \Psi(u - h,v) + \Psi(u,v - h) - \Psi(u-h,v-h)\quad + \nonumber \\ 
    & \frac{h^2}{8}\left[\Psi(u - h,v)\times M^2V_\mathrm{eff}(u - h,v) + \Psi(u,v-h)\times M^2V_\mathrm{eff}(u,v-h)\right]~.
\end{align} 
We now start with initial data on hypersurfaces with $u=$ constant and $v=$ constant. From~\ref{eq:Psi_evol} it is therefore implied that a `forward' point in the evolution is determined completely (and exclusively) by points already evaluated. This means that \ref{eq:RW_master} is self-contained and no additional input condition is necessary for its numerical evolution, which turns out to be its main advantage. In contrast, the evolution using physical coordinates $t,r_*$ requires additional input in the form of boundary conditions on every spatial hypersurface of evolution. Then we evolve our system with Gaussian initial data in $u$ and constant data in $v$. This assumption is a usual choice in previous literature. We show the evolution of perturbations for our system with in~\ref{fig:Echo1} wherein we plot the perturbations with relativistic and non-relativistic DM profiles alongside the case with no DM. The reflecting boundary conditions imply echoes which are a commonly found feature of ECOs. The time period between successive echoes is known as the echo time and is a characteristic feature of the perturbation. As expected from the potentials in~\ref{fig:Pots}, the echoes in the relativistic case are not visibly distinguishable in their features with respect to the DM-less case. However non-relativistic DM produces a different barrier, and thus its echo structure is clearly distinguishable in~\ref{fig:Echo1}. We remember that the echo time  $T_{\rm echo}$ is approximately two times the light travel time between the reflecting surface and the potential barrier in the tortoise coordinate $r_*$. In other words 
\begin{equation}\label{eq:echo_time}
    T_{\rm echo}(M, \zeta) = 2M \left[\zeta - 2(1 + \epsilon + \mathrm{ln}\epsilon)\right],  
\end{equation}
where $r_*=\zeta M$ represents the location of the peak of the effective potential of the object in question. Thus for the relativistic and Schwarschild cases, $\zeta= 3 - 2{\rm log}(2)$ while for the non-relativistic case we see from~\ref{fig:Pots} that $\zeta\approx 100$. So a longer echo time is observed from the non-relativistic case in~\ref{fig:Echo1}. In this way we demonstrate that the computation of GW-echoes in theory can distinguish between different DM structures around ECOs.

\section{Discussion and Conclusion}\label{sec:conclusion}
It is generally agreed upon that environments do have an effect upon different astrophysical systems, but the explicit calculation of such effects has gained prominence only recently in the literature. In this work, we have performed a set of such computations where we investigate the effects of the back reaction of a DM halo upon the spacetime of an ECO. Under the assumption of spherical symmetry, we have computed quantities that are observationally relevant for future ground and space-based GW missions. 
\bigskip

\noindent To begin with we have derived explicit expressions for the Love number which is an observationally relevant parameter in GW-inspirals for both Axial and Polar perturbations and for both the relativistic and non-relativistic cases. We have established that just like DM dressed BHs their ECO counterparts will have Love numbers that inherit an ambiguity from the choice of a relevant length-scale. Therefore, GW observations and subsequent parameter estimations of such composite systems are necessary to fix these uncertainties. Then we also the ECO-DM Love numbers themselves respect the basic sense of the perturbation in the fact that they have a DM independent part and a DM dependent part at linear order in the perturbation. We also note interestingly that the Love numbers contain no undetermined coefficients even though the perturbation variables both the `0th' and first orders in $k$ are arbitrary to the extent of a multiplicative constant. This is a powerful result which tells us that any indeterminacy in the Love numbers from multiplicative constants occurs only at second order in $k$ and hence cannot bias the parameter estimation of ECO-DM systems. Then we consider truncated halos, wherein we find that their exterior spacetime is exactly the same as that of NSs. However we also establish that owing to much larger Love numbers these objects are easily distinguishable from NSs from GWs alone. They cannot pose any threat to the well-established parameter estimation pipelines to infer the NS EoS.  

\bigskip

\noindent We also study the GW-echoes from ECO-DM systems for Axial perturbations. Our results are promising in that they demonstrate that echoes can be used to distinguish between DM scenarios. But we realise our results are limited in their accuracy and scope from computational resources and hence are well-suited only as a `proof of concept' demonstration. Nevertheless given that the echoing wavefunctions show differences, a followup study involving a Bayesian inference of echoing waveforms will be worth attempting. Such a study will be able to distinguish different DM models individually from the vacuum case. We leave such exercises for a future attempt.   

\section*{Acknowledgement}
The authors acknowledge helpful discussions with Sumanta Chakraborty, Sudipta Sarkar, and Paolo Pani.  K.C. acknowledges full funding at FZU from the PPLZ (10005320/0501) project unitl June, 2025 and from the GAČR (24-13079S) project during August, 2025 respectively. The authors acknowledge the use of the PHOEBE cluster at CEICO-FZU.

\appendix
\labelformat{section}{Appendix #1} 
\labelformat{subsection}{Appendix #1}

\section{The source terms}\label{sec:source_term}
\subsection{Axial Case}\label{ssec:sta}
The $\ell = 2$ axial source terms of \ref{eq:odedecomp1} have the following coefficients given by
\begin{align}\label{eq:ax-terms}
    \alpha(\ell=2,x) &= 0 ~,\nonumber \\
    \lambda(\ell=2,x) &= -\frac{6x-2}{x^2(x-1)}~, \nonumber \\
    \gamma(\ell=2,x) &= \frac{4\pi x^2 (4M^2\rho)}{\left(x-\frac{m(x)}{M}\right)}~,\nonumber \\
    \beta(\ell=2,x) &= \frac{4\left(\frac{m(x)}{M}-1\right)}{x(x-1)\left(x - \frac{m(x)}{M}\right)} + \frac{8\pi x[4M^2(\rho + 2p_t)]}{\left(x-\frac{m(x)}{M}\right)}~.
\end{align}

\subsection{Polar Case}\label{ssec:stp}
The $\ell=2$ polar terms of \ref{eq:odedecomp1} have coefficients given by
\begin{align}
    \alpha(\ell=2,x) &= \frac{2 (x-1/2)}{x(x-1)} ~,\nonumber \\
    \lambda(\ell=2,x) &= -\frac{6 x(x-1)+1}{x^2 (x-1)^2}~, \nonumber \\
    \gamma(\ell=2,x) &=- \frac{\left(\frac{m(x)}{M}-1\right)}{2 (x-\frac{m(x)}{M})(x-1)}-  \frac{8 \pi x^{2} (4M^2\rho)}{\frac{m(x)}{ M}}~,\nonumber \\
    \beta(\ell=2,x) &=\frac{6\bigg(\frac{m(x)}{M}-1\bigg)}{x(x-1)\bigg(x-\frac{m(x)}{M}\bigg)}+\frac{\bigg(\frac{m(x)}{M}-1\bigg)\bigg(\frac{m(x)}{M}+1- \frac{2m(x)}{Mx}\bigg)}{(x-1)^{2}\bigg(x-\frac{m(x)}{M}\bigg)^2}\nonumber\\&- \frac{8 \pi  (4M^2\rho)\bigg(2 M^{2}x^{2}-2m(x)x M+m(x)^{2}\bigg)}{4M^2\bigg(x- \frac{m(x)}{M}\bigg)^2}\bigg(\frac{M x}{m(x)}\bigg)~.
\end{align}


\bibliography{ref}
\bibliographystyle{./utphys1}

\end{document}